\documentclass[11pt,twoside]{article}
\usepackage{hepro5}
\usepackage{graphicx}

\usepackage[T1]{fontenc} 

\usepackage{latexsym}
\usepackage{verbatim}

\begin{document}

\vskip 1.0cm
\markboth{M.~Berton et al.}{Parent population of beamed NLS1s}
\pagestyle{myheadings}

\vspace*{0.5cm}
\title{Exploring the parent population of beamed NLS1s: \\
from the black hole to the jet}

\author{M.~Berton$^1$, L.~Foschini$^2$, S.~Ciroi$^1$, A.~Caccianiga$^2$, V.~Cracco$^1$, G.~La Mura$^1$, F.~Di Mille$^3$, 
M.~L.~Lister$^4$, S.~Mathur$^5$, B.~M.~Peterson$^5$, J.~L.~Richards$^4$, E. Congiu$^1$, M. Frezzato$^1$, P.~Rafanelli$^1$}
\affil{$^1$University of Padova, vicolo dell'osservatorio 3, 35122, Padova (Italy)\\
$^2$INAF - Osservatorio astronomico di Brera, via E. Bianchi 46, 23807, Merate (Italy)\\
$^3$Las Campanas Observatory - Carnegie Institution of Washington, Colina el Pino Casilla 601, La Serena (Chile)\\
$^4$Purdue University, 525 Northwestern Avenue, West Lafayette, IN 47907, USA \\
$^5$Ohio State University, 140 West 18th Avenue, Columbus, OH 43210, USA}

\begin{abstract}
The aim of this work is to understand the nature of the parent population of beamed narrow-line Seyfert 1 galaxies (NLS1s), by studying the physical properties of three parent candidates samples: steep-spectrum radio-loud NLS1s, radio-quiet NLS1s and disk-hosted radio-galaxies. In particular, we focused on the black hole mass and Eddington ratio distribution and on the interactions between the jet and the narrow-line region. 
\end{abstract}

\section{Introduction}
Narrow-line Seyfert 1 galaxies (NLS1s) are a class of active galactic nuclei (AGN) that exhibits some peculiar characteristics. They are tipically classified according to their low full width at half maximum (FWHM) of H$\beta$, below 2000 km s$^{-1}$, and their ratio [O III]/H$\beta <$ 3 (Osterbrock \& Pogge 1985). Along with the presence of strong Fe II multiplets in the optical spectra, such characteristics reveal that these AGN are type 1 sources, where the broad line region (BLR) is directly visible, and with a relatively low black hole mass (10$^6$-10$^8$ M$_\odot$), which explains the narrowness of the permitted lines. This "small" black hole, along with the high Eddington ratio characterising these sources (Boroson \& Green 1992), led many to think that NLS1s might be young objects still growing (Mathur 2000). \par
7\% of NLS1s are radio-loud (RLNLS1s, Komossa et al. 2006), and some of them exhibit some interesting blazar-like properties, such as high brightness temperature and a flat or even inverted radio spectrum (Yuan et al. 2008). In 2009 the Fermi satellite detected $\gamma$-ray emission coming from one of these sources, revealing the presence of a relativistic beamed jet (Abdo et al. 2009a). To date, the known $\gamma$-ray emitting NLS1s are 10, and their number is still increasing (e.g. see Yao et al. 2015). This population of flat-spectrum radio-loud NLS1s (F-NLS1s) was extensively studied in Foschini et al. (2015). An important result of their work is that F-NLS1s are likely the low mass tail of $\gamma$-ray AGN, as shown in their Fig.~4. Both of the two blazar classes, BL Lacertae objects and flat-spectrum radio quasars (FSRQs), have indeed a much larger mass, and only FSRQs have an Eddington ratio comparable to that of RLNLS1s. \par
Once the existence of a beamed population is confirmed, the most obvious following step is to understand the nature of its parent population, hence how do these beamed sources look like when randomly oriented. Assuming a typical bulk Lorentz factor $\sim$10 (Abdo et al. 2009b), 10 $\gamma$-ray emitting sources should correspond to more than 2000 misaligned sources (Urry \& Padovani 1995). \par
The first candidates as parent sources are NLS1s with jets viewed at large angles, with a steep radio-spectrum (S-NLS1s) and possibly with an extended radio emission. A few examples can be found in the literature (Gliozzi et al. 2010, Doi et al. 2012, Richards \& Lister 2015), but the currently known sources of this kind are anyway quite rare and definitely less than 2000. A fraction of the parent population is therefore missing. An explanation for this is that the high number of flat-spectrum sources with respect to steep-spectrum sources is due to the high brightness induced by the relativistic beaming: F-NLS1s are more common because we can observe them at larger distances. But this is not the only viable option, because there might be more parent sources that appear as completely different objects (Foschini 2011). \par
In particular, F-NLS1s have a really compact jet and a lack of extended structures at radio frequencies (Foschini et al. 2010) and, when the inclination increases, this jet might become almost invisible. The source would then appear as a radio-quiet NLS1 (RQNLS1). In fact RQNLS1s show a non-thermal radio emission (Giroletti \& Panessa 2009), and a bunch of them even harbor a jet (Doi et al. 2013, 2015). Another possibility is based on a different assumption regarding the BLR geometry. If the BLR is disk-like shaped (Decarli et al. 2008, Shen \& Ho 2014), when observed pole-on the lines are not broadened by the rotation and appear as narrow. When the inclination increases, the Doppler effect broadens the lines, and the galaxy appears as a broad-line radio-galaxy (BLRG). Increasing again the inclination, the line of sight intercepts the molecular torus, and the galaxy is a narrow-line radio-galaxy (NLRG). Moreover, since many studies revealed that NLS1s are always hosted by disk galaxies (Crenshaw et al. 2003), also radio-galaxies must be disk-hosted (disk RGs). This work is a summary of the researches already fully presented in Berton et al. (2015) and Berton et al. (submitted).

\section{Black hole mass}
\begin{figure}  
\begin{center}
\includegraphics[height=8cm,width=10cm]{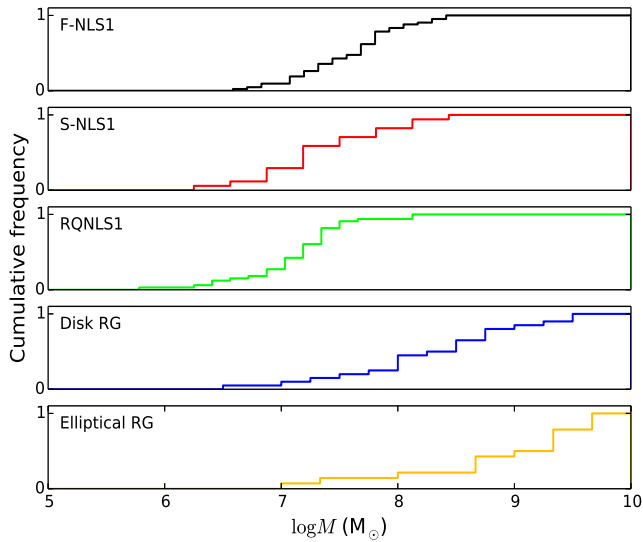}
\caption{{\small Cumulative distributions of black hole masses. From top to bottom, F-NLS1s, S-NLS1s, RQNLS1s, disk RGs and elliptical RGs.}}
\label{mass_dist}
\end{center}
\end{figure}
To understand which one of the previous candidates are actually part of the parent population, we built three samples of candidates, to compare their black hole masses and their accretion luminosities with the sample of F-NLS1s of Foschini et al. (2015), and with a control sample of elliptical radio-galaxies derived from the 2 Jy sample of Inskip et al. (2010). We analysed only sources whose optical spectrum could be obtained from the literature or with the Asiago 1.22m telescope. \par
To calculate the black hole mass we derived the second order momentum $\sigma$ of H$\beta$ and the stellar velocity dispersion from the core component of the [O III] $\lambda$ 5007 line (Nelson \& Whittle 1996, Greene \& Ho 2005). We decided to use line luminosities instead of continuum luminosity, which is often used to calculate the black hole mass, because lines are less affected by the jet contribution. Moreover, we used $\sigma$ instead of FWHM, because the former is less affected by inclination and BLR geometry (Collin et al. 2006). In this way we obtained the mass distributions shown in Fig.~\ref{mass_dist}. NLS1s are all located between 10$^6$ and 10$^8$ M$_\odot$. F-NLS1s have on average a slightly larger mass than the other two NLS1s samples. Disk RGs span instead over a large interval of masses, while elliptical radio-galaxies are located in the high mass region, above 10$^9$ M$_\odot$, as expected. \par
To compare these distributions, we used the Kolmogorov-Smirnov test (K-S), which tests whether two cumulative distributions can be drawn from the same population. An important $-$ and expected $-$ result is that the distributions of steep- and flat-spectrum NLS1s are quite close to each other. This can be interpreted as a sign that these are the same kind of source observed under different angles. The K-S instead cannot provide a conclusive outcome for disk RGs and RQNLS1s. Disk RGs are in the middle between F-NLS1s and elliptical RGs, connecting them with a sort of "bridge". In particular, disk RGs with low black hole mass and high Eddington ratio are well overlapped with the F-NLS1s distribution, suggesting a possible connection between them. We also found that, in presence of a pseudobulge, this connection gets even stronger. Finally, to better understand the role of RQNLS1s, we decided to approach them from a different perspective. 

\section{[O III] lines and narrow-line region}
\begin{figure}[t!]  
\begin{center}
\includegraphics[height=8cm,width=10cm]{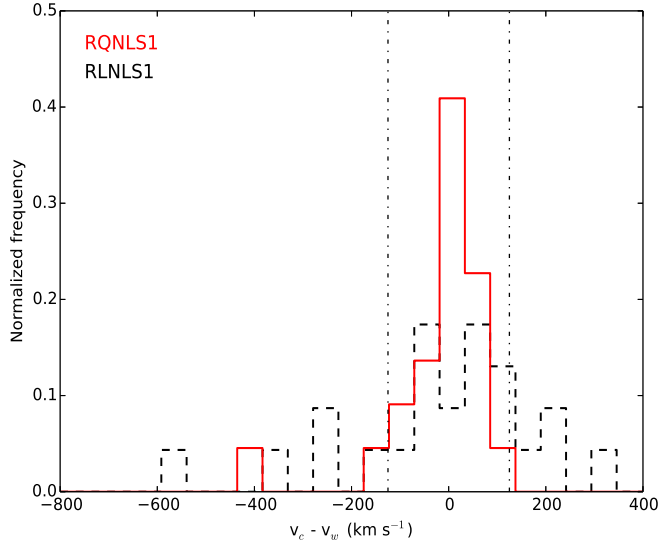}
\caption{{\small Normalized histogram of the [O III] lines shift with respect to their restframe wavelength. The reference used is narrow H$\beta$. Dashed black line is RLNLS1s, solid red line is RQNLS1s. Vertical dash-dotted lines are the limits for blue and red outliers, as in Komossa et al. (2008).}}
\label{blue_out}
\end{center}
\end{figure}
[O III] $\lambda\lambda$ 4959,5007 are forbidden lines originated in the narrow-line region (NLR). These lines are typically the brightest high-ionization lines in the optical spectrum. For this reason, they are particularly suitable for studying their kinematic components. [O III] lines often show an outflowing component within the NLR, called the blue wing. In some peculiar sources they also show a blue-shift of both lines. When this velocity shift is larger than 150 km s$^{-1}$, such sources are known as blue outliers. They are associated with a bulk motion of the NLR toward the observer, possibly induced by an intense radiation pressure coming from the accretion disk or by interaction with a relativistic jet (Zamanov et al. 2002, Komossa et al. 2008). \par
In our work we analysed the [O III] lines of two samples of NLS1s, one radio-loud including both flat- and steep-spectrum, drawn from a sample of very radio-loud sources selected in Yuan et al. (2008), and one radio-quiet, the same used in Berton et al. (2015) for black hole mass calculation. We used two Gaussians for each line, and measured the velocities of core and blue wings components with respect to the narrow component of H$\beta$. One of our results is shown in Fig.~\ref{blue_out}. The histogram represents the velocity shift of the [O III] lines with respect to their restframe wavelength. The distribution of RLNLS1s is wider than that of RQNLS1s ($\sigma_{RL}$ = 193 km s$^{-1}$, $\sigma_{RQ}$ = 103 km s$^{-1}$). This indicates that the [O III] velocities in RLNLS1s can reach higher values than in RQNLS1s. \par
The number of blue outliers is also higher in radio-loud sources (9 against 2), and the K-S test reveals that the two distribution in Fig.~\ref{blue_out} do not arise from the same population (p-value 0.04). We suggest that this might be a sign of an ongoing interaction between the relativistic jet and the NLR clouds in RLNLS1s. In RQNLS1s the NLR is also perturbed, but in a less evident way. This might reveal that, even if some kind of perturbative mechanism is in action, it is something different from, and less effective than, a relativistic jet, and hence that jets may not be tipically harbored in RQNLS1s. \par
Nevertheless, in spite of their classification as radio-quiet sources, non-thermal emission and jets can be observed in a small fraction of RQNLS1s. There are, however, two possible explanations for this apparent contraddiction. The first one is that radio-loudness is not an absolute parameter, because it strongly depends on the way optical and radio fluxes are measured (Ho \& Peng 2001). A radio-quiet classification then does not automatically exclude the presence of a jet. A second possibility is that the jet activity in some NLS1s is intermittent, as suggested in Foschini et al. (2015), and the radio-quiet condition is then only temporary. \par

\section{Future work}
Our previous studies revealed that only steep-spectrum RLNS1s and some low mass disk RGs are likely to be included in the parent population. The numerical problem we presented in the introduction is though still open. Is the number of parent sources high enough to constitute the entire parent population? \par
A very important part of this question can be answered by means of radio luminosity function. The latter is the volumetric density of sources as a function of their radio luminosity. As shown in several works, the great advantage of luminosity functions is that the relativistic beaming can be analytically added to a misaligned function to obtain the observed function of the beamed population (Urry \& Shafer, 1984, Urry \& Padovani, 1991). This allows to directly compare the two populations, and find a solution for the numerical problem (Urry \& Padovani 1995). \par
We therefore created four complete samples of sources, one for F-NLS1s, and three for parent candidates, calculating their luminosity and evaluating the LFs. We also compared our beamed population with a sample of compact steep-spectrum sources (O'Dea 1998), usually believed to be a class of young radio-galaxies often associated with NLS1s (e.g. Caccianiga et al. 2014). Our results will be presented in the upcoming paper by Berton et al. (in prep.). Finally, to deepen the investigation of the radio-quiet/radio-loud relation in NLS1s, we are carrying out dedicated surveys with VLBA and VLA, that will help us to understand the jet launching mechanism and the origin of this dichotomy. 

\acknowledgments \small This research has made use of the NASA/IPAC Extragalactic Database (NED) which is operated by the Jet Propulsion Laboratory, California Institute of Technology, under contract with the National Aeronautics and Space Administration. We acknowledge the usage of the HyperLeda database (http://leda.univ-lyon1.fr). Funding for the Sloan Digital Sky Survey has been provided by the Alfred P. Sloan Foundation, and the U.S. Department of Energy Office of Science. The SDSS web site is \texttt{http://www.sdss.org}. SDSS-III is managed by the Astrophysical Research Consortium for the Participating Institutions of the SDSS-III Collaboration including the University of Arizona, the Brazilian Participation Group, Brookhaven National Laboratory, Carnegie Mellon University, University of Florida, the French Participation Group, the German Participation Group, Harvard University, the Instituto de Astrofisica de Canarias, the Michigan State/Notre Dame/JINA Participation Group, Johns Hopkins University, Lawrence Berkeley National Laboratory, Max Planck Institute for Astrophysics, Max Planck Institute for Extraterrestrial Physics, New Mexico State University, University of Portsmouth, Princeton University, the Spanish Participation Group, University of Tokyo, University of Utah, Vanderbilt University, University of Virginia, University of Washington, and Yale University. This research has made use of the SIMBAD database, operated at CDS, Strasbourg, France.

\end{document}